%% file: O3_tdk.tex
\documentclass[aps,prb,reprint,groupedaddress,showpacs,showkeys,amsmath]{revtex4-1}
\usepackage{graphicx}
\usepackage{dcolumn}
\usepackage{bm}

\begin{document}

%\title{Potential energy surface of the ozone molecule by Resonating Valence Bond Quantum Monte Carlo}
%\title{Resonating Valence Bond Quantum Monte Carlo to compute the potential energy surface of the ozone molecule}
\title{Resonating Valence Bond Quantum Monte Carlo: Application to the ozone molecule}
\author{Sam Azadi}
\affiliation{Department of Physics, Imperial College London, Exhibition Road,
London SW7 2AZ, United Kingdom}
\author{Ranber Singh}
\affiliation{Johannes Gutenberg University Mainz, Institute of Physical Chemistry and Center for Computational Sciences, Staudinger Weg 7, D-55128 Mainz, Germany}
\author{Thoms D. K\"{u}hne*}
\affiliation{Department of Chemistry and Institute for Lightweight Design with Hybrid Systems, University of Paderborn, 
Warburger Str. 100, D-33102 Padeborn, Germany}
\email{tdkuehne@mail.upb.de}
\date{\today}

\input{0_Abstract_tdk.tex}

\maketitle

\input{1_Introduction_tdk.tex}

\input{2_Computational_Details_tdk.tex}

\input{3_Results_tdk.tex}

\input{4_Conclusion_tdk.tex}

\end{document}

%% file: 0_Abstract_tdk.tex
\begin{abstract}
We study the potential energy surface of the ozone molecule by means of Quantum Monte Carlo simulations based on the resonating valence bond concept. The trial wave function consists of an antisymmetrized geminal power arranged in a single-determinant that is multiplied by a Jastrow correlation factor. Whereas the determinantal part incorporates static correlation effects, the augmented real-space correlation factor accounts for the dynamics electron correlation. 
%... accurate ansatz for quantum chemical calculations based on a simple variational wave function, defined by a single geminal expanded in molecular orbitals and combined with a real space correlation factor. 
The accuracy of this approach is demonstrated by computing the potential energy surface for the ozone molecule in three vibrational states: symmetric, asymmetric and scissoring. We find that the employed wave function provides a detailed description of rather strongly-correlated multi-reference systems, which is in quantitative agreement with experiment. 
%We show that our simple but accurate wave function, defined by a single geminal that couples all the electrons into singlet pairs, provides an accurate description of the $O_3$(\~{X}) molecule chemical bond. Our wave function, even in the presence of pseudopotential, yields a large fraction of dynamical and static electron correlation. We obtain almost exact dissociation energy compared with experimental results.   
\end{abstract}

%% file: 1_Introduction_tdk.tex
\section{Introduction}

Beside its environmental and chemical importance, ozone plays a significant role in assessing the accuracy of electronic structure methods due to its multi-reference character. The latter is due to static electron correlation that arises in situations with degeneracy or near-degeneracy, as in transition metal chemistry and strongly-correlated systems in general \cite{Yang}. As a consequence, determining the electronic structure of ozone requires an accurate treatment of both, dynamic as well as static electron correlation effects. Hence, computing the total energy, 
equilibrium geometry and vibrational frequencies of the ozone molecule is a rigorous test for even the most accurate quantum-chemical approaches, such as quantum Monte Carlo (QMC) \cite{Luchow, Lu1, Lu2}, configuration interaction \cite{Shih, Hay, Thunemann, Lee, Raghavachari}, coupled-cluster \cite{Stanton, Magers, Lipscomb, Scuseria, Scuseria2, Watts, Helgaker, Watts2, Bak, Boese, Hino, Grein, Pabst}, as well as multi-reference-based methods \cite{Adler-Golden, Banichevich, Borowski, Barysz, Banichevich2, Banichevich3, Szalay, Borowski2, Li, Grimme, Hubac, Grimme2, Evangelista}, just to name a few.

%In this work, we provide a description of the chemical bonding and electronic structure of the ozone molecule by studying its potential energy surface (PES). In the Born-Oppenheimer approximation, a PES is the relationship between the ground state energy of a molecule and its geometry. An accurate description of the chemical bond plays a central role in obtaining the PES.
In this work, we study the potential energy surface (PES) of a single $O_3$ molecule using the resonating valence bond (RVB) theory of the chemical bond, which was first introduced by Pauling in the early days of quantum mechanics \cite{Pauling1, Pauling2}. According to the RVB theory, a spin singlet can be formed between any two valence electrons belonging to neighboring atoms. Within this approach, the ground state wave function of a molecule can lower the variational energy by imposing an auxiliary resonating degree of freedom between all valence bond configurations. However, the number of bonds increases exponentially with the number of atoms. 
%Fortunately, the molecular orbital (MO) scheme is a strong tool to tackle this problem.  
Nevertheless, it was realized that a single-determinant wave function combined with a suitable real space correlation term, known as that Jastrow correlation factor \cite{Jastrow}, can be employed to represent a complex RVB state \cite{RVB, Bouchaud}. To that extend, the variational energy is computed by means of conventional QMC techniques \cite{QMCrev, Luchow, Austin}, such as variational Monte Carlo (VMC) and diffusion Monte Carlo (DMC) \cite{VMC, DMC}. The RVB wave function is expanded in terms of molecular orbitals (MOs) and optimized by a constrained energy minimization at constant number of MOs \cite{marchi1, marchi2, marchi3}. 
%The most relevant correlations are obtained by using an appropriate number $n$ of MOs. The appropriate value of $n$ is determined by the condition of having size-consistent results based on the atomization to Hartree-Fock Jastrow Slater wave function. The ground state energy and other physical and chemical properties of the system are then obtained by an efficient and accurate variational approach based on QMC techniques. 
In this way, an accurate RVB wave function based on a single-determinant is obtained, which allows to study systems with sizable static and dynamic electron correlation \cite{sam13, sam14}. 
%This allows us to give an accurate description of the chemical bond even in problematic systems where strong dynamical correlations and weak van der Waals interaction are crucial \cite{casula1, casula2}. 
%In fact, QMC-based methods are extremely accurate for taking into account many-body interactions in systems where the accuracy of 1~meV/atom is required \cite{sam13, sam14}. 
%In this work, we present an almost exact dissociation energy and an accurate potential energy surface for the ozone molecule by applying QMC methods together with the RVB wave function. 

%% file: 2_Computational_Details_tdk.tex
\section{Resonating Valence Bond Quantum Monte Carlo}
%Regarding the RVB concept, we consider the electronic wave function of an electron pair. 
The RVB wave function, which is referred to as Jastrow antisymmetrized geminal power (J-AGP), is given by a product of a Jastrow correlation factor $J$ and a determinatal part $\Phi_N$, i.e. $\Psi_{\text{J-AGP}} = J \Phi_N$ \cite{casula1, casula2}. Each geminal is described by a symmetric pairing-function $\phi(\bm{r}_i, \bm{r}_j) =\phi(\bm{r}_j, \bm{r}_i)$. 
%The singlet pair is  described by a symmetric $\phi(\vec r,\vec r^\prime) =\phi(\vec r^\prime,\vec r)$ orbital function. 
For an $N$-electron system, consisting of $N^\uparrow$ spin-up and $N^\downarrow$ spin-down electrons and assuming that $N^\downarrow \leq N^\uparrow$, we can define a pure spin state with a total spin $S=|N^\uparrow-N^\downarrow|/2$ and maximum spin projection $S=S^{tot}_z$ in terms of an antisymmetrized product of 
%even for an arbitrary number of electrons $N=N_\uparrow + N_\downarrow$, 
%where, for the sake of simplicity, we consider here $N_\uparrow\ge N_\downarrow$, $N_\uparrow$ ($N_\downarrow$) being the number of spin up (spin down) electrons.To this end, we take 
$N^\downarrow$ singlet pairs and $2S$ unpaired orbitals $\phi_j(\bm{r}_j^\uparrow)$ of the remaining spin-up electrons. 
%we apply the antisymmetrization operator ${\cal A}$ to obtain a consistent fermionic wave function: 
The resulting antisymmetrized geminal power (AGP) wave function \cite{AGP}, which can be arranged in a single-determinant \cite{RVB, Bouchaud}, reads as 
%Hence, the determinantal part of the J-AGP wave function reads as 
\begin{equation} \label{AGP}
\Phi_N(\bm{R})= {\cal A }  \prod\limits_{i=1}^{N^\downarrow} \phi (\bm{r}^\uparrow_i, \bm{r}^\downarrow_i ) \prod_{j=N^\downarrow+1}^{N^\uparrow} 
\phi_j (\bm{r}^\uparrow_j ),
\end{equation}
where $\bm{R}= \left\{ \bm{r}_1^\uparrow, \cdots, \bm{r}_{N^\uparrow}^\uparrow, \bm{r}_1^\downarrow, \cdots,  \bm{r}_{N^\downarrow}^\downarrow \right\}$ indicates the $3N$ dimensional vector of all electron coordinates, while $\cal A$ is the antisymmetrization operator. 
%The wave function in Eq.~\ref{AGP} is also called the antisymmetrized geminal product (AGP).
In general, breaking number-symmetry leads to the Bardeen-Cooper-Schrieffer (BCS) wave function \cite{BCS}, whereas breaking spin-symmetry yields unrestricted Hartree-Fock determinants \cite{GKLchan}. This is to say that the RVB state corresponds to the Gutzwiller-projected BCS state, while the AGP is the particle-conserving version of the BCS wave function, which accounts for the static electron correlation. 

%Since the spin projection $S_z^{tot}$ of this wave function is the maximum possible compatible with $N_\downarrow$ singlet pairs and $2S$ unpaired spin-1/2 electrons, the total spin of this wave function is definite, namely $S=S_z^{tot}=|N^\uparrow-N_\downarrow|/2$.

%Another important ingredient In the RVB ansatz is the Jastrow correlation factor that allows the suppression of configurations with overlapping valence bonds, which are too expensive in energy and lead to a too large electron density around an atom. 
Dynamic correlation effects between the electrons is taken into account by the Jastrow correlation factor 
%Indeed, the Jastrow term takes into account the electron-electron repulsion and the form usually adopted reads as follows:
\begin{equation}\label{jastrow}
J(\bm{R}) = exp \left(  \sum \limits_{i<j} f ( \bm{r}_i, \bm{r}_j) \right), 
\end{equation}
where $f(\bm{r}_i, \bm{r}_j)$ is a two-electron coordinate function just as the previously introduced pairing-function $\phi(\bm{r}_i, \bm{r}_J)$ and is chosen such as to satisfy the Kato cusp conditions \cite{Kato}.
%is the Jastrow correlation factor, a function of two electron coordinates as the pairing function $\phi$.

Just like any other function of two coordinates, $f(\bm{r}_i, \bm{r}_j)$ as well as $\phi(\bm{r}_i, \bm{r}_j)$ can be expressed in terms of single-particle orbitals: 
\begin{equation} \label{geminal}
\phi( \bm{r}_i, \bm{r}_j) = \sum\limits_{a,b}^{M} \sum\limits_{\mu, \nu}^{m} \lambda^{a,b}_{\mu, \nu} \varphi^{a}_{\mu}( \bm{r}_i) \varphi^{b}_{\nu}( \bm{r}_j), 
\end{equation}
where $\varphi(\bm{r})$ is an atomic orbital (AO), while $m$ is the number of AOs per atom and $M$ the number of atoms. The coefficients $\lambda^{a,b}_{\mu, \nu}$ are arranged in a quadratic $mM \times mM$ matrix denoted as $\bm{\Lambda}$. Moreover, to conserve the total spin, $\bm{\Lambda}$ is also symmetric. Assuming that the AOs are non-orthogonal, i.e. $S^{a,b}_{\mu, \nu}=\langle \varphi | \varphi \rangle \neq \bm{I}$, diagonalizing $\bm{\Lambda}$ corresponds to solving the equation 
\begin{equation} \label{diag}
  \bm{\Lambda SP} = \bm{P\Lambda}^\prime, 
\end{equation}
where the matrix $\bm{P}$ contains the eigenvectors of $\bm{\Lambda}$ and diagonal matrix $\bm{\Lambda}^\prime$ the associated eigenvalues $|\lambda_1^\prime| \ge |\lambda_2^\prime| \ge \cdots |\lambda_{mM}^\prime| \ge 0$. The number of nonzero eigenvalues is equivalent to the rank of $\bm{\Lambda}$. Using the identity $\bm{P}^{T}\bm{SP} = \bm{I}$ that is identical to $\bm{P}^{T}=(\bm{SP})^{-1}$ and multiplying it to the right of Eq.~(\ref{diag}), $\bm{\Lambda}=\bm{P\Lambda}^{\prime}\bm{P}^{T}$ is obtained. From this it follows that Eq.~(\ref{geminal}) is equivalent to 
\begin{equation} \label{pairing}
\phi( \bm{r}_i, \bm{r}_j) = \sum\limits_{k=1}^{n-2S} \lambda_k^\prime \phi_k( \bm{r}_i) \phi_k( \bm{r}_j) 
\end{equation}
where $n$ is the number of MOs $\phi_k( \bm{r}_i)$ that can be written as a linear combination of AOs: 
\begin{equation} \label{LCAO}
\phi_k( \bm{r}_i) = \sum\limits_{l} c_{kl} \varphi_l(\bm{r}_i), 
\end{equation}
%where $n$ is large enough, and $\{ \phi_j \}$ is an orthogonal single particle basis set, that becomes complete in the CBS (complete basis set) limit for $n\to \infty$ \cite{sam10}.
where $c_{kl}$ are the AO coefficients that independently on the particular basis set used are implicitly dependent on $\bm{r}_i$. 
%
%Notice also that in these notations we assume that the $2S$ unpaired orbitals 
%$\phi_j$ correspond to the indices: $ n-2S+1 \le j \le n$.

%We can particularly choose the MO obtained by a conventional HF calculation for $\phi_j$ and truncate the expansion to the number of MOs $n$ equal to the number of electron pairs and unpaired orbitals, namely $n=N_\uparrow$. 
The smallest number of orbitals, which is compatible with an unpolarized $N$-electron system, is $n=N^\uparrow$. In that case the rank of $\bm{\Lambda}$ is also minimal and identical to $n$ and the antisymmetrization operator ${\cal A}$ singles out just one Slater determinant (SD). 
%In this case, we recover the usual HF theory with just $N_\uparrow$ MO, because the antisymmetrization operator ${\cal A}$ singles out just one Slater determinant (SD).
Due to the fact that the determinantal part of the resulting wave function is equivalent to the one of unrestricted Hartree-Fock (HF) theory, we will refer to it as JHF to emphasize the presence of the Jastrow correlation factor. 
%As such, the RVB wave function can be thought of as the natural extension of the HF theory, where the Jastrow correlation factor accounts for the dynamic electron correlation, while for $n> N^{\uparrow}$ most of the static correlation is recovered. 
 %Moreover, the MO weights $\lambda_j$ only affect an overall prefactor of this Slater determinant, so that their actual values are irrelevant. 
If the orbitals of the SD are determined by means of density functional theory (DFT), the eventual wave function is denoted as JDFT \cite{sam10}.
However, the great appeal of the AGP wave function is that if the rank of $\bm{\Lambda}$ is larger than $N^{\uparrow}$, it corresponds to a multireference wave function that is able to describe static electron correlation effects, although at the computational cost of a genuine single-determinant approach. We will name the resulting wave function as JAGP. 
%The pairing function is generally not limited to having only $N^\downarrow$ non-vanishing eigenvalues $\lambda_j$. In general, a gain in energy and a more accurate calculation are expected whenever $n> N^{\uparrow}$. More details of the method can be found at Ref.\cite{marchi1}.  
%The most relevant correlations are obtained by using an appropriate number $n$ of MOs. The appropriate value of $n$ is determined by the condition of having size-consistent results based on the atomization to Hartree-Fock Jastrow Slater wave function. The ground state energy and other physical and chemical properties of the system are then obtained by an efficient and accurate variational approach based on QMC techniques. 
As such, the RVB wave function can be thought of as the natural extension of the HF theory, where the Jastrow correlation factor accounts for the dynamic electron correlation, while for $n> N^{\uparrow}$ most of the static correlation is recovered. 
%, while determinantal part includes the HF wave function in a particular limit, and goes beyond HF when $n>N_\uparrow$.

%% file: 3_Results_tdk.tex
\section{Computational details}

All of our QMC calculations were performed using the {\it TurboRVB} program package of Sorella and coworkers \cite{turbo}. We have employed the 
two different trial wave functions, the JDFT and the JAGP wave function. In both cases the single-determinant is made of MOs as obtained by DFT within local-density approximation (LDA), as described in Ref.~\onlinecite{sam10}. In case of the JAGP wave function all parameters were variationally optimized at the VMC level of theory using the stochastic reconfiguration algorithm \cite{SR, benzene, Hdimer}, while for the JDFT wave function the parameters of the Jastrow factor were held fixed. 
%The determinant part is a pairing wave function which describes the correlation between electron pairs using geminal function, as explained in the previous section.
Scalar-relativistic energy consistent pseudopotentials \cite{pseudo} were used to describe two core electrons of the oxygen atoms. For the single-determinant we employed an atomic basis set consisting of contracted $10s,8p,6d,4f$ Gaussian-type orbitals, while for the Jastrow correlation factor uncontracted $2s,2p,1d$ Gaussian-type orbitals were used. The variational energies were computed using two different QMC methods, VMC and the even more accurate lattice-regularised diffusion Monte Carlo (LRDMC) projection technique \cite{lrdmc}, which eventually results in four trial wave functions we have investigated here: JDFT-VMC, JDFT-LRDMC, JAGP-VMC and JAGP-LRDMC. %Combining the wave-functions employed and QMC methods, we compared the results obtained by JDFT-VMC, JDFT-LRDMC, JAGP-VMC and JAGP-LRDMC schemes. 

\section{Results and discussion}

The PES of the $O_3$ molecule exhibits two types of minima, one of $C_{2v}$ symmetry (open minimum) with an apex angle of $116.75^\circ$ and another one of $D_{3h}$ symmetry (ring minimum), where the oxygen atoms form an equilateral triangle. Due to the fact that the open structure is energetically significantly lower than the closed ring case, we will confine ourselves to the $O_3$ molecule with $C_{2v}$ symmetry only. Symmetry considerations suggest the existence of three open minima, which are separated by energetically high barriers and thus can be considered as independent from each other \cite{siebert}. As a consequence, the potential energy $E(d_1, d_2, \alpha)$ is calculated as a function of just three variables, where $d_1$ and $d_2$ are the two $O_c-O$ bond distances, while $\alpha$ is the $O-O_{c}-O$ bond angle and $O_{c}$ the central atom.
%PES is calculated with respect to just three variables: the two $O_c-O$ bond distances $d_1$ and $d_2$, as well as the $O-O_{c}-O$ bond angle $\alpha$, where $O_{c}$ is the central atom. 
The experimental values for $d_1$, $d_2$, $\alpha$, and the dissociation energy are $2.4052 a_0$, $2.4052 a_0$, $116.75^\circ$, and $26.105~kcal/mol$, respectively \cite{EXP}. %In In our results, the experimental zero point energy correction \cite{EXP} is included. 

%\begin{video}
% \includegraphics{sym.gif}
% \caption{\label{symmetryvideo} symmetrical vibration}
%\end{video}
\begin{figure}
    \centering
    \includegraphics[width=0.475\textwidth]{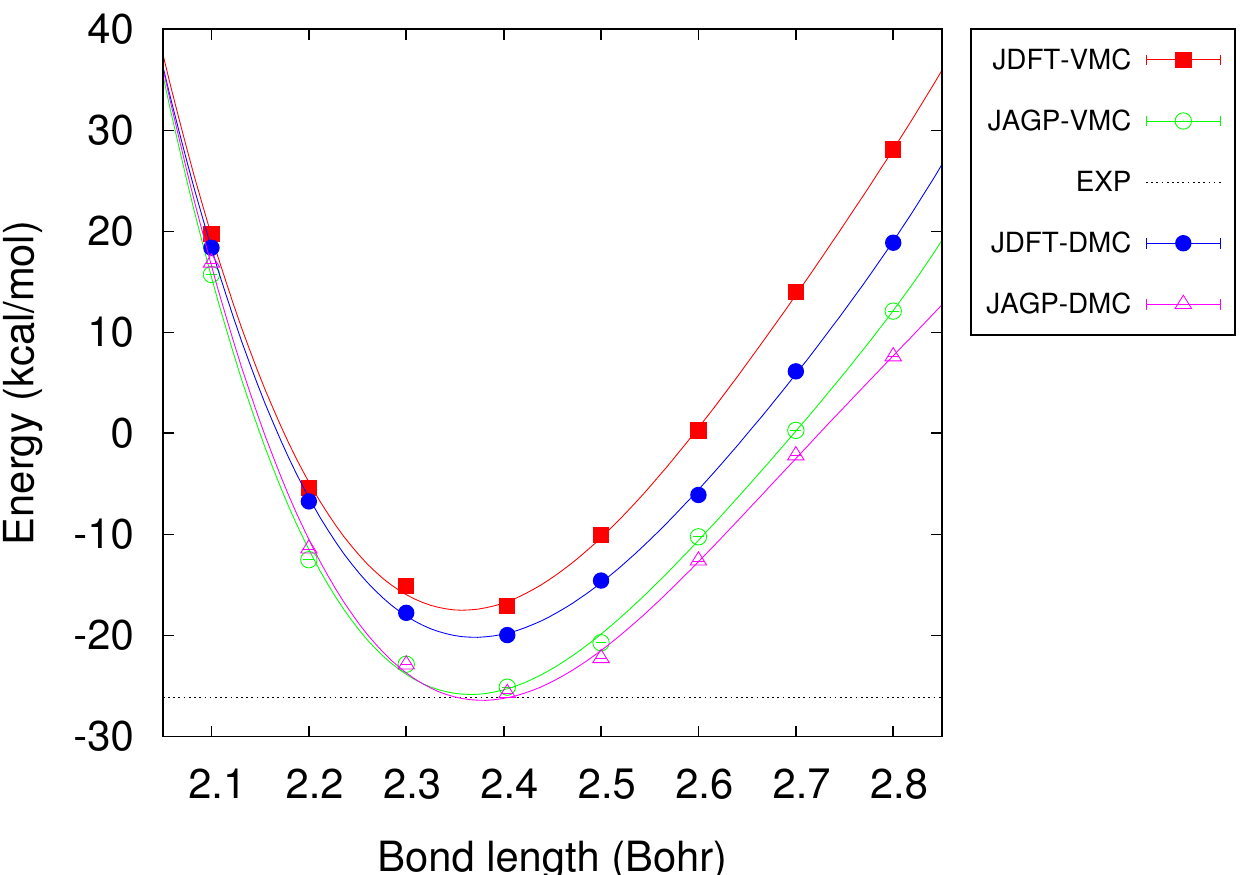}
    \caption{\label{O2Osym}(colour online) The PES of the $O_3$(\~{X}) molecule in a symmetrical vibration state as obtained by the JDFT-VMC, JDFT-LRDMC, JAGP-VMC and JAGP-LRDMC methods. The energy normalization is such that zero energy corresponds to the ground state energy of $O_2(^3\Sigma_{g}^-)+O(^3P)$. The experimental dissociation energy is illustrated by the dotted line.}
\end{figure}
Figure~\ref{O2Osym} illustrates the PES of the $O_3$(\~{X}) molecule in a symmetrical vibration state,  %obtained by JDFT-VMC, JDFT-LRDMC, JAGP-VMC and JAGP-LRDMC methods. The zero energy reference corresponds to a total ground state energy of $O_2(^3\Sigma_{g}^-)+O(^3P)$. In the symmetrical vibration state, 
where the value of $\alpha$ is fixed at $116.8^\circ$ and the $d_1=d_2$. 
%Therefore, instead of $d_1$ and $d_2$, we just use bond length as the independent variable. 
The compressed $O_3$(\~{X}) molecule is dominated by Coulomb repulsion and the results of all trial wave functions investigated here rather similar. 
%In the positive dissociation energy regime, JAGP and JDFT wave functions provide similar results, regardless of the QMC methods used. By decreasing the both bond lengths and bringing oxygen together, JAGP-WF tends to JDFT-WF. While the potential energy surfaces obtained by different wave functions and QMC methods separate significantly in the bonding area, 
However, whenever the $O_3$(\~{X}) molecule is stretched, static electron correlation becomes more important and the JAGP wave function superior due to the strong multi-reference character of the stretched $O-O$ bond. The dissociation energies at the experimental equilibrium geometry, using the JDFT-VMC, JDFT-LRDMC, JAGP-VMC, as well as JAGP-LRDMC methods are -16.60(5), -19.74(4), -25.25(6) and -26.14(5)~kcal/mol, respectively. 
%In the case of JDFT-WF, switching from VMC to LRDMC improves the dissociation energy by $19\%$. This switching only improves the dissociation energy by $3.5\%$ if the JAGP-WF is employed. 
In fact, even in at the variational level, the JAGP wave function outperforms accurate LRDMC calculations based on the JDFT trial wave function and recovers $\sim 97\%$ of the experimental dissociation energy, while at the JAGP-LRDMC level the result is correct up to $0.1\%$.

\begin{figure}
    \centering
    \includegraphics[width=0.475\textwidth]{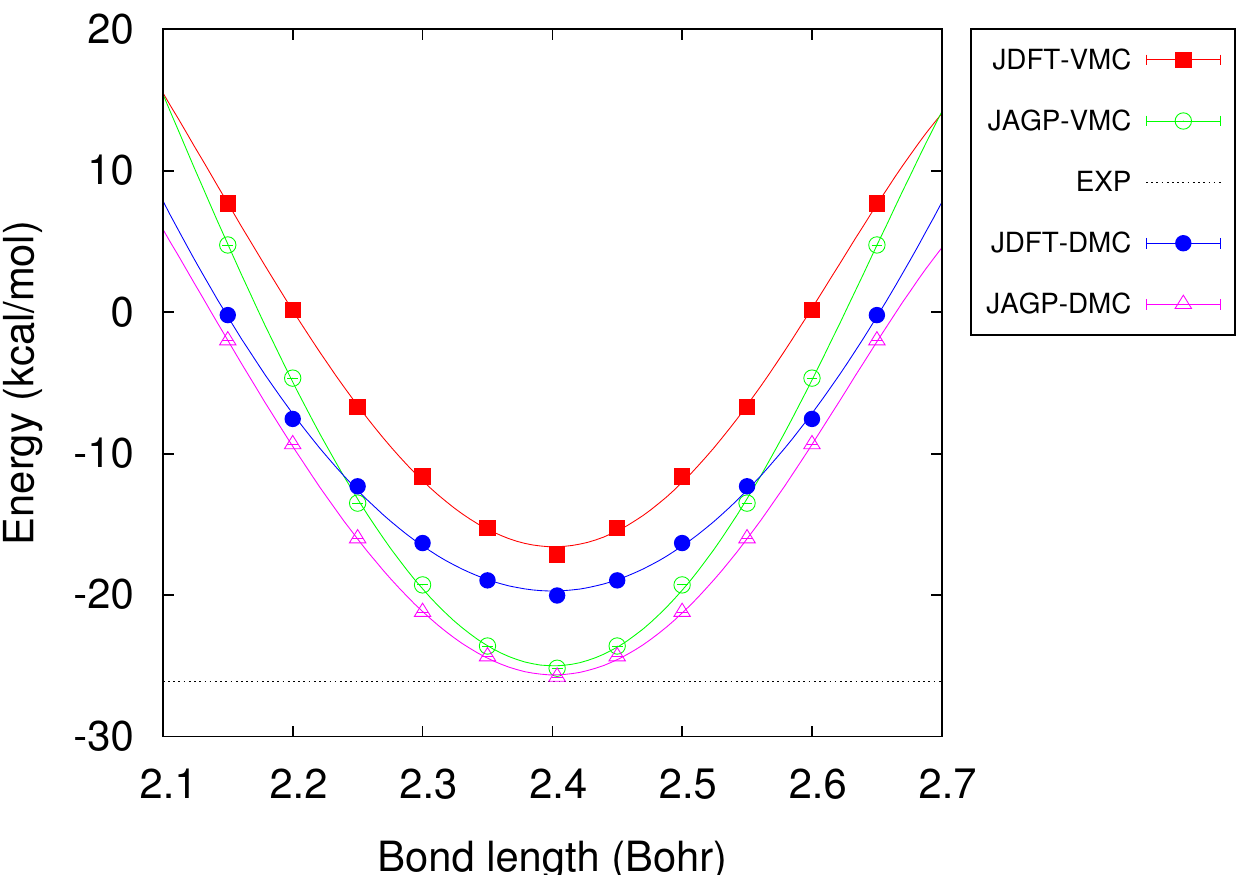}
    \caption{\label{O2Oasym}(colour online) The PES of the $O_3$(\~{X}) molecule in an asymmetrical vibration state as obtained by the JDFT-VMC, JAGP-VMC, JDFT-LRDMC and JAGP-LRDMC methods. The energy normalization is such that zero energy corresponds to the ground state energy of $O_2(^3\Sigma_{g}^-)+O(^3P)$. The experimental dissociation energy is illustrated by the dotted line.}
\end{figure}
The PES of the $O_3$(\~{X}) molecule in an asymmetrical stretching state is shown in Figure~\ref{O2Oasym}, where 
%obtained by JDFT-VMC, JDFT-LRDMC, JAGP-VMC and JAGP-LRDMC methods. The zero energy reference corresponds to total ground state energy of $O_2(^3\Sigma_{g}^-)+O(^3P)$.
the value of $\alpha$ is again fixed at $116.8^\circ$. Defining $\delta d$ as deviation from the equilibrium bond length, each point on the PES corresponds to $d_1 \pm \delta d$ and $d_2 \mp \delta d$. Given the symmetry, $E (d_1+\delta d, d_2-\delta d, \alpha) = E (d_1-\delta d, d_2+\delta d, \alpha)$, as can been seen in the axial symmetry around the equilibrium bond length in Figure~\ref{O2Oasym}. The corresponding dissociation energies for the JDFT-VMC, JDFT-LRDMC, JAGP-VMC, as well as JAGP-LRDMC techniques are -16.58(6), -19.71(5), -24.99(7) and -25.64(5)~kcal/mol, respectively. %The dissociation energy obtained using JAGP-WF and LRDMC methods is up to $98.2\%$ of the experimental one. For $0 < \delta d < 1.5$ (Bohr), the JAGP-VMC produces lower energy than JDFT-LRDMC. 
Interestingly, at equilibrium the JAGP is significantly superior to the JDFT trail wave function regardless of the particular QMC method, whereas for large asymmetries the LRDMC technique is essential. 
%Therefore, within bonding distance, JAGP-WF provides a much more accurate description of bonding compared to JDFT-WF, even when the computation method is improved. It has to be noticed that the LRDMC method generally substantially improves the variational energy \cite{benzene}. 

\begin{figure}
    \centering
    \includegraphics[width=0.475\textwidth]{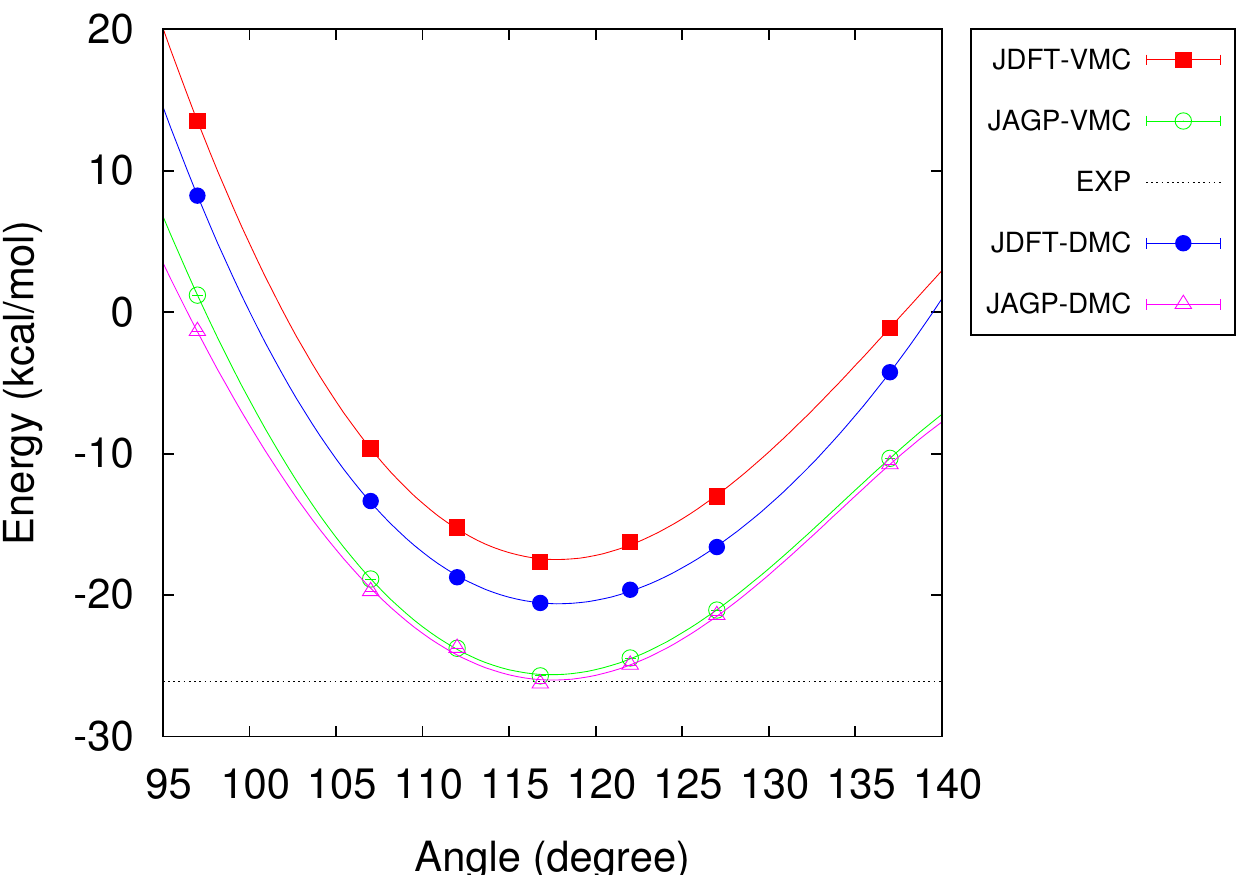}
    \caption{\label{O2Oben}(colour online) The PES of the $O_3$(\~{X}) molecule in a scissoring vibration state as obtained by the JDFT-VMC, JDFT-LRDMC, JAGP-VMC and JAGP-LRDMC methods. The energy normalization is such that zero energy corresponds to the ground state energy of $O_2(^3\Sigma_{g}^-)+O(^3P)$. The experimental dissociation energy is illustrated by the dotted line.}
\end{figure}
Figure ~\ref{O2Oben} exhibits the PES of the $O_3$(\~{X}) molecule in a scissoring state, where 
%obtained by JDFT-VMC, JDFT-LRDMC, JAGP-VMC and JAGP-LRDMC methods. The zero energy reference corresponds to a total ground state energy of $O_2(^3\Sigma_{g}^-)+O(^3P)$. 
$d_1$ and $d_2$ are both fixed at $2.4052 a_0$ and, thus, $\alpha$ the only variable. At the equilibrium bond angle, the dissociation energies as obtained by the JDFT-VMC, JDFT-LRDMC, JAGP-VMC and JAGP-LRDMC methods are -17.42(7), -20.55(5), -25.61(6), and -25.97(5)~kcal/mol, respectively. Again, the JAGP systematically improves the description of the JDFT wave function. In fact, even at the JAGP-VMC method exceeds JDFT-based LRDMC calculations. Moreover, the small difference between the JAGP-VMC and JAGP-LRDMC schemes clearly demonstrates accuracy of the JAGP approach. 
%The difference between dissociation energies obtained using JAGP-WF but different QMC methods is only 0.37 kcal/mol indicating the accuracy of JAGP-WF to describe the chemical bond, even at the variational level. 

%% file: 4_Conclusion_tdk.tex
\section{Conclusions} 

%In summary, we investigated the potential energy surface of the $O_3$(\~{X}) molecule. We employed two different wave functions: JDFT and JAGP. Total ground state energies were obtained utilising QMC-based VMC and LRDMC methods. The PES of the $O_3$(\~{X}) molecule with $C_{2v}$ symmetry was obtained within three vibrational states: symmetrical, asymmetrical and scissoring. In the case of the symmetrical state and using JAGP-LRDMC we obtained almost exact dissociation energy -26.14(5) kcal/mol, compared to the experimental one which is -26.105 kcal/mol. The dissociation energy corresponding to the asymmetrical vibrational state PES calculated by using JAGP-LRDMC was -25.64(5) kcal/mol, which is up to $98.2\%$ of the experimental results. In the case of the scissoring vibrational state, the difference between our AGP-LRDMC dissociation energy and the experimental one was only 0.135 kcal/mol. Our results clearly show the accuracy of JAGP wave function to describe the chemical bond, even at the variational level. 
In summary, we find that the JAGP systematically improves upon the conventional JDFT wave function in all cases we have considered here. In particular, except for the most asymmetric configurations, the JAGP-VMC method was able to outperform even more accurate LRDMC calculations based on the JDFT trail wave function. Together with the fact that the differences between the JAGP-VMC and JAGP-LRDMC results were rather small, this is a clear manifestation of the superiority of the JAGP wave function. 
%The fact that for the JAGP wave function the differences between the VMC and LRDMC were rather small 
In any case, the significant deviation between the JDFT and JAGP trail wave functions highlights that for a multi-reference system, such as the $O_3$(\~{X}) molecule, an accurate treatment of static electron correlation effects is essential. 
%In conclusion we would like to emphasize two points we find important: firstly, the importance of the optimization procedure, and secondly, the accuracy of JAGP wave function for describing the chemical bond. In fact, the very remarkable outcome of the accurate optimization of JAGP wave function is that it is possible to describe the chemical bond well in most of the interesting oxygen-based compounds. It has to be mentioned that the JAGP wave function usually contains a large number of non-linear variational parameters, which are quite problematic for the optimization. The occurrence of local minima in the potential energy surface leads to the complex numerical problem of finding the universal minimum energy. In addition, because of a strong dependence between several variational parameters, the variation of some nonlinear parameters in the wave function can sometimes be highly compensated by a corresponding deviation of other parameters. However, the considerable difference between the JDFT wave function results and those obtained using JAGP wave function clearly indicates the importance of static correlation in this system. According to our results, the accurate optimization of JAGP wave function parameters plays a crucial role in calculating the static correlation. 

%% file: O3_tdk.bbl
\begin{thebibliography}{99}

\bibitem{Yang} A. J. Cohen, P. Mori-Sanchez and W. Yang, Science {\bf 321}, 792 (2008)
\bibitem{Luchow} S. Manten and A. L\"uchow, J. Chem. Phys. {\bf 115}, 5362 (2001).
\bibitem{Lu1} S.-I. Lu J. Chem. Phys. {\bf 120}, 14 (2004).
\bibitem{Lu2} S.-I. Lu, J. Chem. Phys. {\bf 120}, 10423 (2004).
\bibitem{Shih} S. Shih, R. J. Buenker and S. D. Peyerimhoff, Chem. Phys. Lett. {\bf 28}, 463 (1974).
\bibitem{Hay} P. J. Hay, T. H. Dunning Jr. and William A. Goddard III, J. Chem. Phys. {\bf 62}, 3912 (1975). 
\bibitem{Thunemann} K.-H. Thunemann, S. D. Peyerimhoff and R. J. Buenker, Chem. Phys. Lett. {\bf 70}, 432 (1978).
\bibitem{Lee}  T. J. Lee, W. D. Allen and H. F. Schaefer III, J. Chem. Phys. {\bf 87}, 7062  (1987). 
\bibitem{Raghavachari} K. Raghavachari, G. W. Trucks, J. A. Pople and E. Replogle, Chem. Phys. Lett. {\bf 158}, 207 (1989). 
%\bibitem{Pople} J. A. Pople, M. Head-Gordon, K. Raghavachari and G. W. Trucks, Chem. Phys. Lett. {\bf 157}, 479 (1989). 
%\bibitem{Scuseria} G. E. Scuseria, T. J. Lee, A. C. Scheiner and H. F. Schaefer III, J. Chem. Phys. {\bf 90}, 5635  (1989). 
\bibitem{Stanton} J. F. Stanton, W. N. Lipscomb, D. H. Magers and R. J. Bartlett, J. Chem. Phys. {\bf 90}, 1077 (1989).
\bibitem{Magers} D. H. Magers, W. N. Lipscomb, R. J. Bartlett and J. F. Stanton, J. Chem. Phys. {\bf 91}, 1945 (1989).
\bibitem{Lipscomb} J. F. Stanton, R. J. Bartlett, D. H. Magers and W. N. Lipscomb, Chem. Phys. Lett. {\bf 163}, 333 (1989).
\bibitem{Scuseria} T. J. Lee and G. E. Scuseria, J. Chem. Phys. {\bf 93}, 489 (1990). 
\bibitem{Scuseria2} G. E. Scuseria and T. J. Lee, J. Chem. Phys. {\bf 93}, 5851 (1990). 
\bibitem{Watts} J. D. Watts, J. F. Stanton and R. J. Bartlett, Chem. Phys. Lett. {\bf 178}, 471 (1991).
\bibitem{Helgaker} T. Helgaker, J. Gauss, P. Jorgensen and J. Olsen, J. Chem. Phys. {\bf 106}, 6430 (1997). 
\bibitem{Watts2} J. D. Watts and R. J. Bartlett, J. Chem. Phys. {\bf 108},  2511 (1998). 
\bibitem{Bak} K. L. Bak, J. Gauss, P. Jorgensen, J. Olsen, T. Helgaker and J. F. Stanton, J. Chem. Phys. {\bf 114}, 6548 (2001). 
\bibitem{Boese} A. D. Boese, M. Oren, O. Atasoylu, J. M. L. Martin, M. K‡llay and J Gauss, J. Chem. Phys. {\bf 120}, 4129 (2004).
\bibitem{Hino} O. Hino, T. Kinoshita, G. K. L. Chan and R.J. Bartlett, J. Chem. Phys. {\bf 124}, 114311 (2006).
\bibitem{Grein} F. Grein, J. Chem. Phys. {\bf 130}, 124118 (2009).
\bibitem{Pabst} M. Pabst, A. Kohn, J. Gauss and J. F. Stanton, Chem. Phys. Lett. {\bf 495}, 135 (2010). 
%\bibitem{Bartlett}  R. J. Bartlett, J. D. Watts, S. A. Kucharski and J. Noga, Chem. Phys. Lett. {\bf 165},  513 (1990).
%\bibitem{Kucharski} S.A. Kucharski, R.J. Bartlett, J. Chem. Phys. {\bf 110},  8233 (1999).
%\bibitem{Kinoshita} T. Kinoshita, O. Hino, R.J. Bartlett, J. Chem. Phys. {\bf 123},  074106 (2005). 
\bibitem{Adler-Golden} S. M. Adler-Golden, S. R. Langhoff, C. W. Bauschlicher and G. D. Carney, J. Chem. Phys. {\bf 83}, 255 (1985).
\bibitem{Banichevich} A. Banichevich, S.D. Peyerimhoff and F. Grein, Chem. Phys. Lett. {\bf 173}, 1 (1990).
\bibitem{Borowski} P. Borowski, K. Andersson, P.-A. Malmqvist and B. O. Roos, J. Chem. Phys. {\bf 97}, 5568 (1992).
\bibitem{Barysz} M. Barysz, M. Rittby and R. J. Bartlett, Chem. Phys. Lett. {\bf 193}, 373 (1992).
\bibitem{Banichevich2} A. Banichevich and S.D. Peyerimhoff, Chem. Phys. Lett. {\bf 174}, 93 (1993).
\bibitem{Banichevich3} A. Banichevich, S.D. Peyerimhoff and F. Grein, Chem. Phys. Lett. {\bf 178}, 155 (1993).
\bibitem{Szalay} P. G. Szalay and R. J. Bartlett, Chem. Phys. Lett. {\bf 214}, 481 (1993).
\bibitem{Borowski2} P. Borowski, M. F\"ulscher, P.-A. Malmqvist and B. O. Roos, Chem. Phys. Lett. {\bf 237},195 (1995).
\bibitem{Li} X. Z. Li and J. Paldus, J. Chem. Phys. {\bf 110}, 2844 (1999).
\bibitem{Grimme} S. Grimme and M. Waletzke, J. Chem. Phys. {\bf 111}, 5645 (1999).
\bibitem{Hubac} I. Hubac, J. Pittner and P. Carsky, J. Chem. Phys. {\bf 112},  8779 (2000). 
\bibitem{Grimme2} S. Grimme, M. Parac and M. Waletzke, Chem. Phys. Lett. {\bf 334}, 99 (2001).
\bibitem{Evangelista} F. A. Evangelista, E. Prochnow, J. Gauss and H. F. Schaefer III, J. Chem. Phys. {\bf 132}, 074107 (2010).
%\bibitem{Knowles} P. J. Knowles and H-J. Werner, Chem. Phys. Lett. {\bf 145},  514 (1988).
%\bibitem{Werner} H-J. Werner and P. J. Knowles, J. Chem. Phys. {\bf 89}, 5803 (1988).
%\bibitem{Yamaguchi} Y. Yamaguchi, M. J. Frisch, T. J. Lee, H. F. Schaefer III and J. S. Binkley, Theor. Chim. Acta {\bf 69}, 337 (1986).
%\bibitem{Pittner} J. Pittner, P. Nachtigall, P. Carsky, J. Masik, J. Chem. Phys. {\bf 110},  10275  (1999). 
%\bibitem{Li2} X.Z. Li, J. Paldus, J. Chem. Phys. {\bf 107},  6257 (1997).
\bibitem{Pauling1} L. Pauling, Nature {\bf 161}, 1019 (1948).
\bibitem{Pauling2} L. Pauling, Proc. Roy. Soc. A {\bf 196}, 343 (1949).
%\bibitem{pauling} L. Pauling, in \emph{The nature of the chemical bond}, Third edition, Cornell University Press, Ithaca, New York, page 204.
\bibitem{Jastrow} R. Jastrow, Phys. Rev. {\bf 98}, 1479 (1955).
\bibitem{RVB} P. W. Anderson, Science {\bf 235}, 1196 (1987).
\bibitem{Bouchaud} J. P. Bouchaud and C. Lhuillier, Europhys. Lett. {\bf 3}, 1273 (1987).
\bibitem{QMCrev} W. M. C. Foulkes, L. Mitas, R. J. Needs and G. Rajagopal, Rev. Mod. Phys. {\bf 73}, 33 (2001).
\bibitem{Luchow} A. L\"uchow, WIREs Comput. Mol. Sci. {\bf 1}, 388 (2011).
\bibitem{Austin} B. M. Austin, D. Y. Zubarev and W. A. Lester, Chem. Rev. {\bf 112}, 263 (2012).
\bibitem{VMC} W. L. McMillan, Phys. Rev. {\bf 138}, A442 (1965).
\bibitem{DMC} D. M. Ceperley and B. J. Alder, Phys. Rev. Lett. {\bf 45}, 566 (1980).
\bibitem{marchi1} M. Marchi, S. Azadi, M. Casula and and S. Sorella,  J. Chem. Phys. {\bf 131}, 154116 (2009).
\bibitem{marchi2} M. Casula, M. Marchi, S. Azadi and S. Sorella, Chem. Phys. Lett. {\bf 477}, 255 (2009).
\bibitem{marchi3} M. Marchi, S. Azadi and S. Sorella, Phys. Rev. Lett. {\bf 107}, 086807 (2011).
\bibitem{sam13} S. Azadi, W. M. C. Foulkes and T. D. K\"uhne, New J. Phys. {\bf 15}, 113005 (2013). 
%\bibitem{Calcavecchia} F. Calcaveccia, F. Pederiva, M. H. Kalos and T. D. K\"uhne, Phys. Rev. E {\bf 90}, 053304 (2014). 
\bibitem{sam14} S. Azadi, B. Monserrat, W. M. C. Foulkes and R. J. Needs, Phys. Rev. Lett. {\bf 112}, 165501 (2014).
\bibitem{casula1} M. Casula and S. Sorella, J. Chem. Phys. {\bf 119}, 6500 (2003).
\bibitem{casula2} M. Casula, C. Attaccalite and S. Sorella, J. Chem. Phys. {\bf 121}, 7110 (2004).
\bibitem{AGP} A. C. Hurley, J. E. Lennard-Jones and J. A. Pople, Proc. R. Soc. A {\bf 220}, 446 (1953).
\bibitem{BCS} J. Bardeen, L. N. Cooper and J. R. Schrieffer, Phys. Rev. {\bf 108}, 1175 (1957).
\bibitem{GKLchan} G. K.-L. Chan, WIREs Comput. Mol. Sci. {\bf 2}, 907 (2012). 
\bibitem{Kato} T. Kato, Comm. Pure Appl. Math. {\bf 10}, 151 (1957).
\bibitem{sam10} S. Azadi, C Cavazzoni, and S. Sorella, Phys. Rev. B {\bf 82}, 125112 (2010).
\bibitem{turbo} S. Sorella, {\it TurboRVB Quantum Monte Carlo package}%, URL \underline{http://people.sissa.it/~sorella/web/index.html}.
\bibitem{SR} S. Sorella, Phys. Rev. B {\bf 71}, 241103(R) (2005).
\bibitem{benzene} S. Sorella, M. Casula and D. Rocca, J. Chem. Phys. {\bf 127}, 014105 (2007).
\bibitem{Hdimer} F. Calcavecchia and T. D. K\"uhne, arXiv:1501.01428 (2015).
\bibitem{pseudo} M. Burkatzki, C. Filippi, and M. Dolg, J. Chem. Phys. {\bf 126}, 234105 (2007).
\bibitem{lrdmc} M. Casula, C. Filippi, and S. Sorella, Phys. Rev. Lett. {\bf 95}, 100201 (2005).
\bibitem{siebert} R. Siebert, P. Fleurat-Lessard, R. Schinke, M. Bittererova, and S. C. Farantos, J. Chem. Phys. {\bf 116}, 9749 (2002). 
\bibitem{EXP} Vl. G. Tyuterev, S. Tashkun, D. W. Schwenke, P. Jensen, T. Cours, A. Barbe, and M. Jacon, Chem. Phys. Lett. {\bf 316}, 271 (2000). 

\end{thebibliography}
